\documentclass[final]{svjour3}
\usepackage{graphicx}
\usepackage{rotating}
\usepackage{amssymb}
\usepackage{mathptmx}
\usepackage[numbers]{natbib}
\usepackage{lineno}
\usepackage{url}
\usepackage{hyperref}
\usepackage{siunitx}
\makeatletter
\journalname{Journal of Low Temperature Physics}

\bibpunct{}{}{,}{s}{}{,}

\begin{document}

\newcommand{\hdblarrow}{H\makebox[0.9ex][l]{$\downdownarrows$}-}
\title{Low Noise Frequency-Domain Multiplexing of TES Bolometers Using Sub-kelvin SQUIDs}

\author{
T.~Elleflot$^1$ \and 
A.~Suzuki$^1$ \and 
K.~Arnold$^2$ \and 
C.~Bebek$^1$ \and 
R.~H.~Cantor$^3$ \and 
K.~T.~Crowley$^4$ \and 
J.~Groh$^5$ \and 
T.~de~Haan$^6$ \and 
A.~Hornsby$^4$ \and 
J.~Joseph$^1$ \and
A.~T.~Lee$^{1,4}$ \and 
T.~Liu$^7$ \and 
J.~Montgomery$^8$ \and 
M.~Russell$^2$ \and 
Q.~Yu$^4$}

\institute{
$^1$Physics Division, Lawrence Berkeley National Laboratory, Berkeley, CA 94720, USA; 
$^2$Department of Physics, University of California San Diego, San Diego, CA 92093, USA; 
$^3$STAR Cryoelectronics, Santa Fe, NM 87508, USA; 
$^4$Department of Physics, University of California Berkeley, Berkeley CA, 94720, USA; 
$^5$ National Institute of Standards and Technology, Boulder, CO 80305, USA; 
$^6$ High Energy Accelerator Research Organization (KEK), Tsukuba, Ibaraki 305-0801, Japan; 
$^7$ Electrical and Computer Engineering Department, Princeton University, Princeton, NJ 08540, USA; 
$^8$ Physics Department, McGill University, Montreal, QC H3A 0G4, Canada\\
\email{tuckerelleflot@gmail.com}}

\maketitle

\begin{abstract}

Digital Frequency-Domain Multiplexing (DfMux) is a technique that uses MHz superconducting resonators and Superconducting Quantum Interference Device (SQUID) arrays to read out sets of Transition Edge Sensors. DfMux has been used by several Cosmic Microwave Background experiments, including most recently POLARBEAR-2 and SPT-3G with multiplexing factors as high as 68, and is the baseline readout technology for the planned satellite mission \emph{LiteBIRD}. Here, we present recent work focused on improving DfMux readout noise, reducing parasitic impedance, and improving sensor operation. We have achieved a substantial reduction in stray impedance by integrating the sensors, resonators, and SQUID array onto a single carrier board operated at 250 $\si{\milli\kelvin}$. This also drastically simplifies the packaging of the cryogenic components and leads to better-controlled crosstalk. We demonstrate a low readout noise level of 8.6 $\si{\pico\ampere}/\si{\hertz}^{1/2}$, which was made possible by operating the SQUID array at a reduced temperature and with a low dynamic impedance. This is a factor of two improvement compared to the achieved readout noise level in currently operating Cosmic Microwave Background experiments using DfMux and represents a critical step toward maturation of the technology for the next generation of instruments. 


\end{abstract}

\vspace{-4 mm}
\section{Introduction}

Digital Frequency-Domain Multiplexing (DfMux) \cite{dobbs2012, bender2014, gottardi2016} is a TES multiplexing technique, in which sets of $\sim$1 $\si{\ohm}$ TES bolometers are placed in series with superconducting LC resonators \cite{rotermund2016} that define unique $\mathcal{O}$(1) $\si{\mega\hertz}$ bias frequencies. A voltage bias is applied to each detector at the resonance frequency of its associated LC resonator and the resulting current is actively nulled. Residual current is amplified by a cryogenic SQUID array and room temperature electronics and used to rapidly update the nulling current \cite{dehaan2012}. DfMux has been deployed in many Cosmic Microwave Background (CMB) experiments, including most recently POLARBEAR-2 and SPT-3G with multiplexing factors as high as 68 \cite{bender2018}. DfMux is also the baseline readout technology for the planned CMB satellite mission \emph{LiteBIRD} \cite{hazumi2020}. 

Here, we present recent work aimed at reducing readout noise and parasitic impedance, and improving readout packaging\cite{dehaan2020, lowitz2020}. In section 2, we describe the hardware used for these measurements, including a technology demonstrator circuit board called the Cold Integrated Multiplexing Module (CIMM). The CIMM moves all cryogenic readout components onto a single circuit board that is cooled to 250 $\si{\milli\kelvin}$, reducing parasitic impedances that generate crosstalk and detector instability and simplifying the readout packaging. In section 3, we present results from measurements performed using the CIMM, which include a factor of two reduction in readout noise and factor of five reduction in parasitic resistance compared to currently operating CMB DfMux systems. We conclude in Section 4.

\begin{figure}[!htbp]
\begin{center}
\includegraphics[width=0.6\linewidth, keepaspectratio]{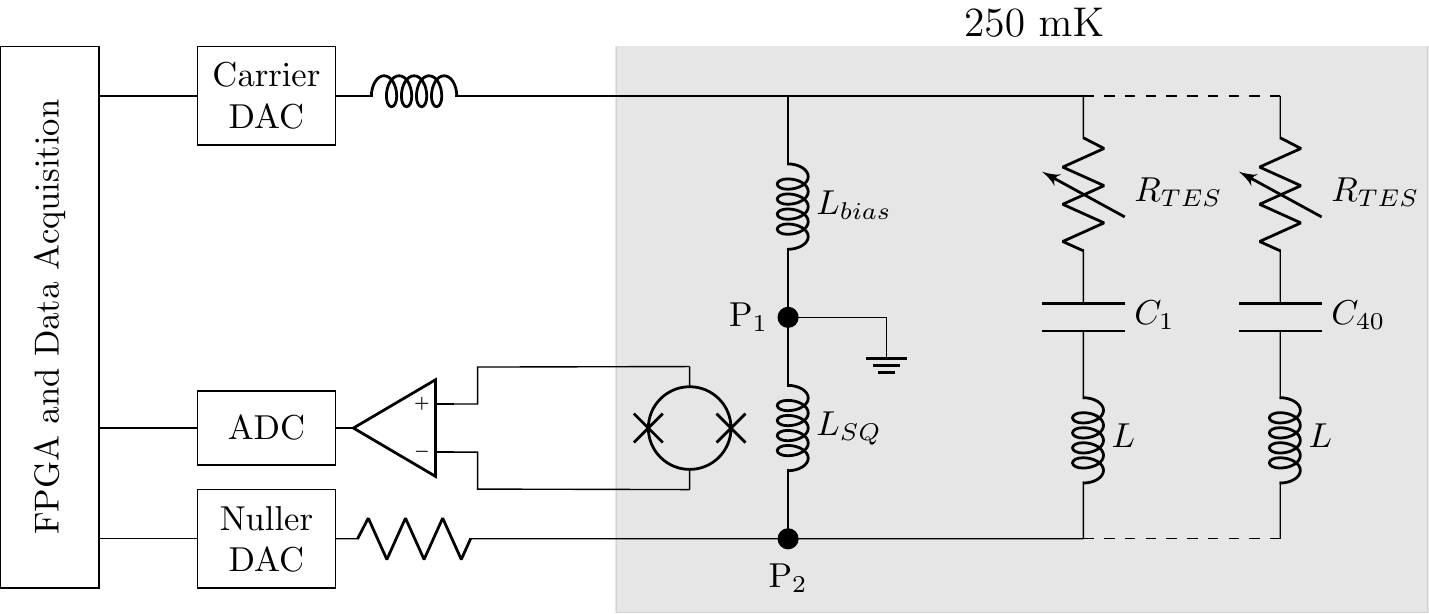}
\includegraphics[width=0.3\linewidth, keepaspectratio]{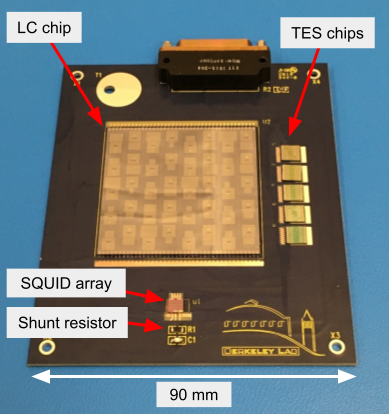}
\end{center}
\caption{(Color figure online) \textit{Left}: Schematic of DfMux circuit with cryogenic components in gray shaded box. Inductance between points \emph{P1} and \emph{P2} is a source of bias frequency-dependent readout noise\cite{bender2018}. We minimize off-chip parasitic inductance by putting points \emph{P1} and \emph{P2} on the SQUID chip itself. \textit{Right}: Photograph of CIMM printed circuit board with readout components labelled.}
\vspace{-4 mm}
\label{fig1}
\end{figure}

\vspace{-4 mm}
\section{Implementation}

This work was done using a technology demonstrator circuit board called the CIMM, described previously in\cite{dehaan2020}. In contrast with deployed Dfmux systems, in which the cryogenic readout components are spread across two cryogenic temperature stages \cite{avva2018}, the CIMM holds all cryogenic readout components on a single printed circuite board at 250 $\si{\milli\kelvin}$. This dramatically reduces the parasitic impedance associated with the electrical connection between the SQUID and resonators, which is a source of crosstalk\cite{dobbs2012, montgomery2021}, and relaxes the requirements on the $\mathcal{O}$(1) $\si{\meter}$ cable that spans the 4 $\si{\kelvin}$ to 250 $\si{\milli\kelvin}$ temperature differential, allowing us to use commercial twisted pair cable instead of the custom low-inductance cables used in deployed instruments. 

Due to a subtlety of the SQUID feedback scheme used, the inductance of the SQUID array input coil and parasitic series inductance provide a mechanism for bias frequency-dependent readout noise\cite{bender2018,montgomery2021}. We designed the CIMM such that only on-chip impedances contribute to this effect (see Fig. 1). We chose to use an SA13ax SQUID array because of its low input inductance of 70 $\si{\nano\henry}$\cite{silvafeaver2018}. The SA13ax has several other favorable characteristics as well, including high transimpedance ($Z_t=dV/dI_{in}$ where $V$ is the voltage across the array and $I_{in}$ is the current through the input coil) at 4 $\si{\kelvin}$ of $Z_t\sim$700 $\si{\ohm}$ and variable dynamic impedance ($R_{dyn}=dV/dI_{SQ}$ where $I_{SQ}$ is the current through the SQUID array). The SA13ax is also used in the SPT-3G and PB-2 receivers at a temperature of 4 $\si{\kelvin}$.

The CIMM also holds one LC resonator chip and up to five TES chips, allowing us to operate 40 detectors. The TES bias is provided by a 4.2 $\si{\nano\henry}$ controlled inductance circuit board trace. The use of a reactive bias element reduces power dissipation on the sub-kelvin stage and provides a small reduction in readout noise.

Magnetic shielding was provided by several layers of high magnetic permeability foil (Metglas 2705M). Care was taken to ensure that no ferromagnetic materials were used in the fabrication of the circuit board. The assembly was installed on the sub-kelvin stage of a cryostat cooled with a pulse tube cryocooler and a helium adsorption cooler with a base temperature of 250 $\si{\milli\kelvin}$. The connection from the CIMM to the room temperature electronics is made with twisted pair cable.

\vspace{-4.2 mm}
\section{Results}

\subsection{SQUID Performance}

We measured flux modulation curves at several temperatures between 4 $\si{\kelvin}$ and 250 $\si{\milli\kelvin}$. The modulation curves were smooth at 4 $\si{\kelvin}$, but irregularities developed on the rising slope of curve at/below 800 $\si{\milli\kelvin}$. The SA13ax array has significantly different dynamic impedance values on the rising and falling slope of the modulation curve, with $R_{dyn}\sim$300 $\si{\ohm}$ on the rising slope $R_{dyn}\sim$700 $\si{\ohm}$ on the falling slope (in the polarity that we have chosen). Operating the array with low $R_{dyn}$ is crucial to achieving the required reduction in readout noise and these irregularities would prevent us from doing so. These irregularities are also thought to be associated with excess SQUID noise. Others have demonstrated reduced SQUID noise by using an RC snubber in parallel with the SQUID input coil \cite{audley2020}. We found that shunting the input coil with a cryogenic resistor completely removed the irregularities at all accessible temperatures (see Fig. 2). The resistor and input inductance of the SQUID input coil form a low-pass filter. We chose the shunt resistance value such that the cutoff frequency of the filter is well above our readout bandwidth. During CIMM operation, the shunt resistor is at a temperature of 250 $\si{\milli\kelvin}$ and its Johnson noise makes a negligible contribution to the total readout noise.

SQUID tuning is the process by which a current bias and flux bias is applied to the SQUID in order to achieve an optimum between readout noise and linearity. Tuning is usually performed using an automated routine, but can also be performed manually. We performed the automated SQUID tuning routine at several temperatures between 4 $\si{\kelvin}$ and 250 $\si{\milli\kelvin}$ and found that the transimpedance, $Z_t=dV/dI_{in}$, and dynamic impedance, $R_{dyn}=dV/dI_{SQUID}$, increased by 60\% and 20\% over this temperature range, respectively. We also investigated the achievable range of $Z_t$ and $R_{dyn}$ values by manually tuning the SQUID to eleven different bias points while holding the CIMM at a constant temperature of 250 $\si{\milli\kelvin}$ (Fig. 4). Using manual tuning, it is possible to increase $Z_t$ by nearly a factor of two relative to the automated routine. But since $R_{dyn}$ is strongly correlated with $Z_t$, the expected improvement in readout noise from manual tuning is less than a factor of two. The impact of SQUID tuning on readout noise will be discussed more in Section 3.3.

\begin{figure}[htbp]
\begin{center}
\includegraphics[width=\linewidth, keepaspectratio]{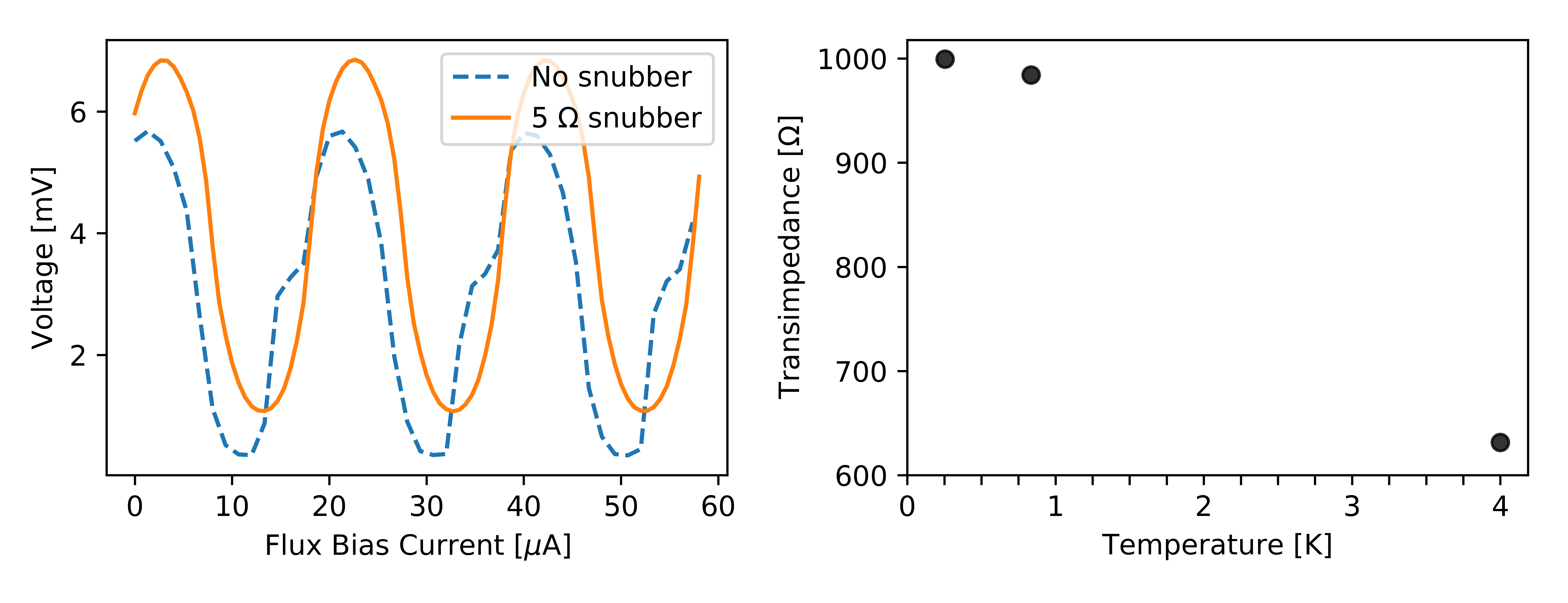}
\end{center}
\caption{(Color figure online) \textit{Left}: The modulation curve for an SA13ax SQUID array measured at 250 $\si{\milli\kelvin}$ with and without a 5 $\si{\ohm}$ shunt resistor in parallel with its input coil. \textit{Right}: Transimpedance of an SA13ax SQUID array as a function of temperature. The array was tuned using an automated tuning algorithm that optimizes between $Z_t$ and linearity.}
\vspace{-4 mm}
\label{fig1}
\end{figure}

\subsection{TES Operation}
We verified TES operation by simultaneously lowering the voltage biases applied to a set of TES bolometers such that they cooled through their superconducting transitions. This was done by first heating the CIMM above the critical temperature of the sensors ($\sim$450 $\si{\milli\kelvin}$) and applying a large voltage bias of $\sim$10 $\si{\micro\volt}$ using the CIMM's inductive bias element. We then cooled the CIMM to its base temperature and then incrementally decreased the voltage biases while measuring the resulting current. The results are shown in Fig 3. A total of 30 bolometers were electrically connected and we successfully operated 28. There are broad stable regions on many channels. Several channels become unstable at high fractional resistance and we suspect that this is due to a mismatch between the detector and electrical bandwidths, rather than instability due to parasitic impedance. The residual resistance measured after the detector is superconducting\cite{elleflot2018, elleflot2020} contributes to crosstalk and detector instability at low operating resistance. The measured values have a median of 60 $\si{\milli\ohm}$, which is a reduction by about a factor of five compared to currently deployed systems.

\begin{figure}[htbp]
\begin{center}
\includegraphics[width=0.8\linewidth, keepaspectratio]{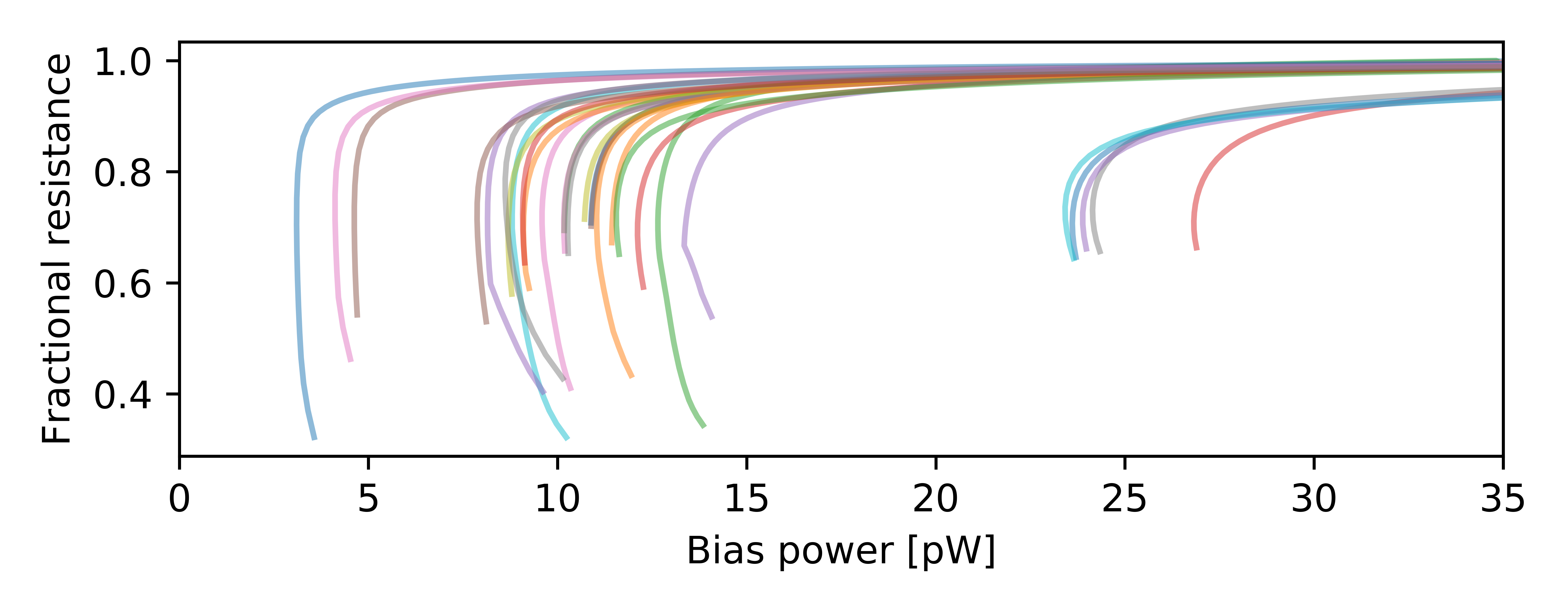}
\end{center}
\caption{(Color figure online) A set of TES bolometers are cooled through their superconducting transition by lowering the applied voltage biases. The TES bolometers used had varied normal resistance and saturation power.}
\vspace{-4 mm}
\label{fig1}
\end{figure}

\subsection{Readout Noise}

Power absorbed by a bolometer is transduced into an amplitude-modulation of the narrow-band SQUID feedback current associated with the bolometer's bias frequency \cite{dehaan2012}. In the absence of signals and detector noise, we can take the demodulated signal as a measurement of readout noise. To perform this measurement, we replaced the TES sensors with 1 $\si{\ohm}$ resistors, which do not have a superconducting transition in the relevant temperature range and do not have sensitivity to optical power. The resistors contribute Johnson noise to the measurement, which we conservatively estimated by assuming a low electron temperature of 250 $\si{\milli\kelvin}$ and a resistance of 1 $\si{\ohm}$. This is removed when analyzing readout noise data, though it only accounts for about 10\% of the total measured noise.

Since the SQUID array used can be operated with low or high $R_{dyn}$, we were able to check the impact of $R_{dyn}$ on readout noise. The SQUID dynamic impedance impacts readout noise in two ways: it converts the input current noise of the first stage amplifier into a voltage noise at the output of the SQUID, and it forms a low-pass filter with the capacitance of the cryogenic wiring between the SQUID and 300 $\si{\kelvin}$ amplifier \cite{montgomery2021}. The latter effect was not expected to be important for this measurement because we are using a short length of cryogenic wiring with capacitance of $\sim$30 pF, resulting in a cutoff frequency of 8 MHz and 16 MHz for high $R_{dyn}$ and low $R_{dyn}$ tuning, respectively, both of which are well above our readout bandwidth. Comparing the readout noise from high and low $R_{dyn}$ SQUID tuning, we measured a noise reduction of 6-8 $\si{\pico\ampere}/\si{\hertz}^{1/2}$ with low $R_{dyn}$ tuning across the range of bias frequencies, which matched our expectations from the current noise of the room temperature amplifier.

\begin{figure}[htbp]
\begin{center}
\includegraphics[width=.8\linewidth, keepaspectratio]{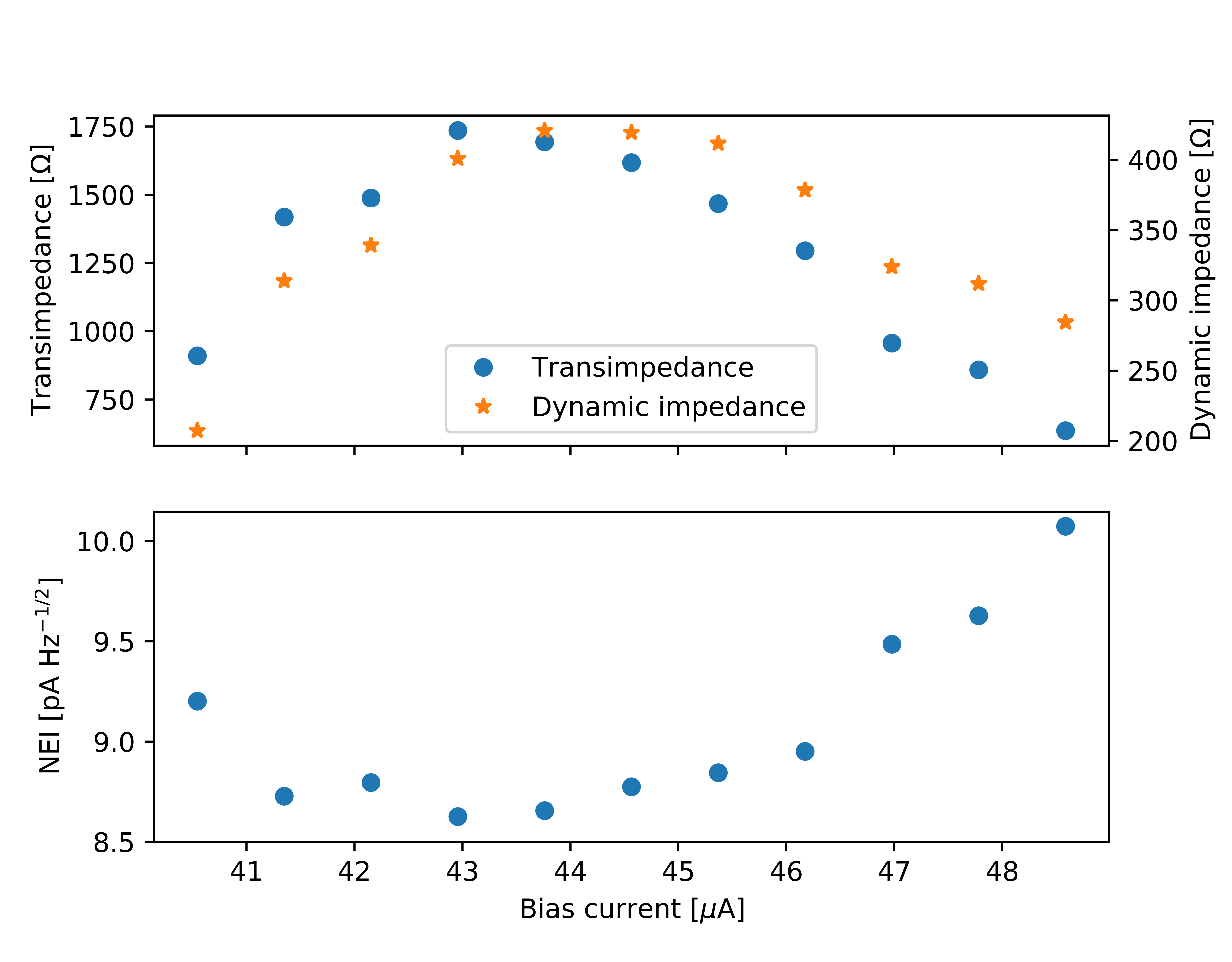}
\end{center}
\caption{(Color figure online) Measured values of $Z_t$ and $R_{dyn}$ (\textit{top}) and readout noise (\textit{bottom}) at 250 $\si{\milli\kelvin}$ with several different values of current bias applied to the SQUID array.}
\vspace{-4 mm}
\end{figure}

Next, we looked at the effect of optimizing SQUID tuning on readout noise. We measured readout noise while the CIMM was cooled to 250 $\si{\milli\kelvin}$ for eleven different manual SQUID tunings. Despite the change in $Z_t$ and $R_{dyn}$ by a factor of about two over the range of current biases, we measure only a $\sim$10\% change in median readout noise. This is likely due to the correlation between $Z_t$ and $R_{dyn}$ and the fact that they decrease and increase readout noise, respectively. There is a broad minimum in readout noise, which indicates that there is flexibility in optimizing SQUID linearity without significantly affecting readout noise performance. The lowest achieved median noise level is 8.6 $\si{\pico\ampere}/\si{\hertz}^{1/2}$. This represents a factor of two reduction in readout noise compared to currently observing CMB experiments using DfMux. The bias frequency dependence of the readout noise is also greatly suppressed. We measure an increase in noise by about 15\% at our highest bias frequencies, which should be compared to a $\sim$100\% increase in noise in deployed CMB DfMux systems.

We refer the measured readout noise to the input of a \emph{LiteBIRD}-like bolometer by assuming a 1 $\si{\ohm}$ TES operating resistance and a conservative 1 $\si{\pico\watt}$ electrical bias power\cite{westbrook2020}. In all observing bands, the measured readout noise is below the projected \emph{LiteBIRD} sensitivity\cite{matsumura2016}.

\begin{figure}[h!]
\centering
\includegraphics[width=0.8\linewidth, keepaspectratio]{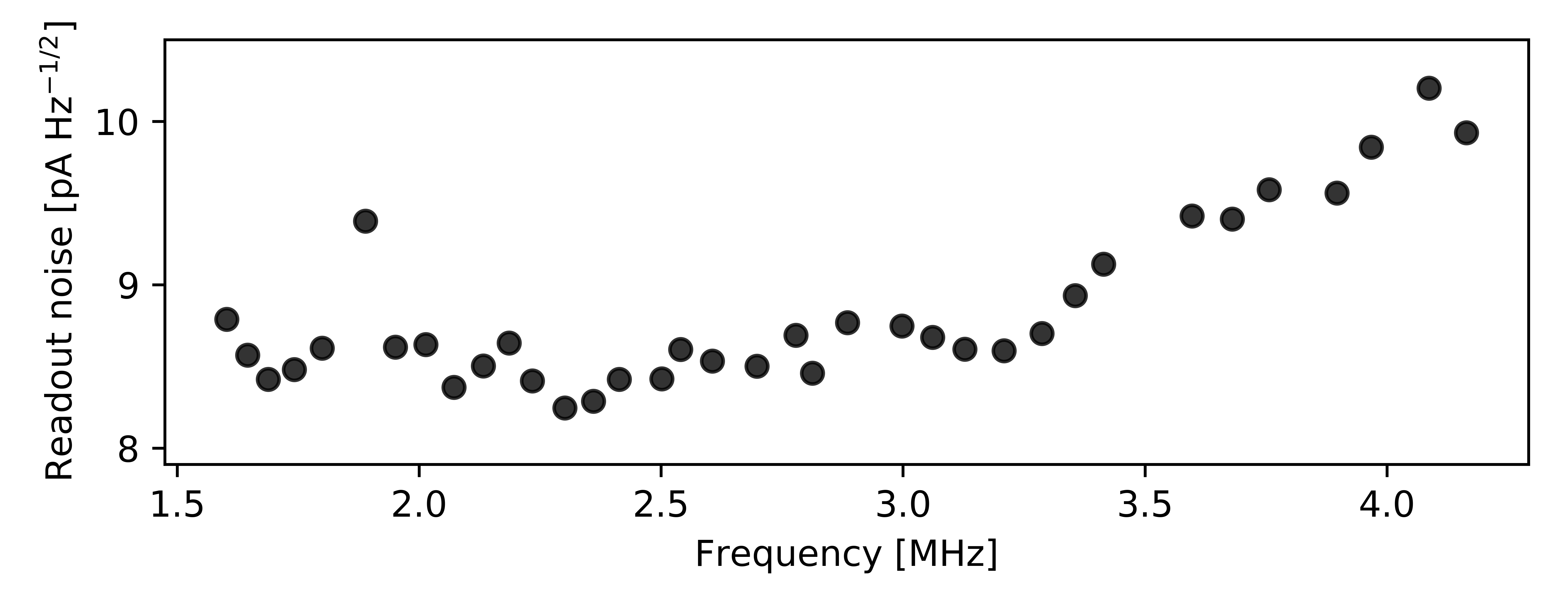}
\caption{Measured readout noise as a function of bias frequency for 36 channels. The median value is 8.6 $\si{\pico\ampere}/\si{\hertz}^{1/2}$. The slight increase in noise with frequency is likely due to an effect related to the SQUID feedback\cite{bender2018} and is suppressed significantly compared to deployed instruments.}
\label{fig1}
\end{figure}

\vspace{-4 mm}
\section{Conclusions}

We have demonstrated several improvements to TES multiplexing using DfMux readout, including a substantial reduction in readout noise and operation of many TES bolometers under an inductive bias. The reduction in readout noise is a crucial step toward future applications of DfMux readout. The noise reduction was achieved by operating the SQUID array at lower temperature and with low dynamic impedance tuning. Additionally, parasitic impedances which are known to affect the frequency dependence on the readout noise have been significantly reduced by the circuit board design. The decrease in residual superconducting resistance in the TES bias circuit will allow for TES operation at lower resistance, resulting in higher responsivity and a further suppression of readout noise compared to deployed CMB DfMux systems. The improvements made to the cryogenic architecture significantly relax the constraints on the cryogenic cables, making the cryogenic packaging more robust and much simpler to procure. We are developing a version of the CIMM that is compatible with the POLARBEAR-2 receivers and we hope to deploy several CIMM modules in the near future. The datasets generated during and/or analysed during the current study are available from the corresponding author on reasonable request.

\begin{acknowledgements}
This work was supported by Early Career Research Program, Office of Science, of the U.S. Department of Energy.
Many of the figures were made using matplotlib \cite{hunter2007} and PGF/TikZ \cite{tikz}. Analyses
were conducted using python scientific packages \cite{perez2007, vanderwalt2011}.
\end{acknowledgements}

\pagebreak

\bibliographystyle{unsrt}
\bibliography{references} 

\end{document}